\documentclass[10pt, conference]{IEEEtran}

\usepackage{amsmath,amssymb}
\usepackage[linesnumbered,ruled]{algorithm2e}
\usepackage{graphicx,xcolor,array}
\usepackage{pgfplots}
\pgfplotsset{compat=1.14}
\usepackage{tikz,cite}
\usetikzlibrary{arrows}
\usetikzlibrary{automata,positioning}
\usepackage{makecell}
\usepackage{upgreek}
\usepackage{pifont}
\usepackage{mathscinet}
\usepackage[letterpaper, left=0.625in,right=0.625in,top=0.75in,bottom=1in]{geometry}
\usepackage{geometry}

\usepackage{array}

\usepackage{cite}
\usepackage{amsmath,amssymb,amsfonts}
\usepackage{algorithmic}
\usepackage{graphicx}
\usepackage{textcomp}
\usepackage{xcolor}
\usepackage{enumitem}
\usepackage{amsthm}

\newtheorem{definition}{Definition}

\usepackage[colorlinks=true,linkcolor=black,anchorcolor=black,citecolor=black,filecolor=black,menucolor=black,runcolor=black,bookmarks=false,hidelinks,urlcolor=black]{hyperref}
\usepackage{enumitem}

\newif\ifcomment
\commenttrue

\makeatletter
\def\endthebibliography{%
  \def\@noitemerr{\@latex@warning{Empty `thebibliography' environment}}%
  \endlist
}
\makeatother

\begin{document}

\title{A Review on Privacy in DAG-Based DLTs}

\author{
\IEEEauthorblockN{
Mayank Raikwar
\IEEEauthorblockA{University of Oslo, Norway\\
Email: mayankr@ifi.uio.no}}}


\maketitle
\begin{abstract} Directed Acyclic Graph (DAG)-based Distributed Ledger Technologies (DLTs) have emerged as a promising solution to the scalability issues inherent in traditional blockchains. However, amidst the focus on scalability, the crucial aspect of privacy within DAG-based DLTs has been largely overlooked. This paper seeks to address this gap by providing a comprehensive examination of privacy notions and challenges within DAG-based DLTs. We delve into potential methodologies to enhance privacy within these systems, while also analyzing the associated hurdles and real-world implementations within state-of-the-art DAG-based DLTs. By exploring these methodologies, we not only illuminate the current landscape of privacy in DAG-based DLTs but also outline future research directions in this evolving field.
    
\end{abstract}

\section{Introduction}

Privacy in distributed ledgers encompasses two key aspects: identity privacy and data privacy, including transaction amounts and user balances. Concepts like transaction unlinkability and untraceability also contribute to privacy considerations. Various privacy-centric cryptocurrency systems, e.g.,~\cite{Monero,Zcash,Zether,Pribank} built on blockchain technology prioritize user and data confidentiality, employing cryptographic techniques for privacy assurance. However, existing DAG-based DLTs primarily prioritize performance enhancement, sidelining user and transaction privacy to avoid complexity that might impact transaction confirmation times. Yet, as cryptographic methods mature, integrating techniques like Zero-knowledge proofs~\cite{sun2021survey}, ring signatures~\cite{sun2017ringct}, and anonymous broadcast channels~\cite{kohlweiss2021anonymity} for identity privacy or transaction mixing, confidential transactions, and encryption (e.g., encrypted mempool~\cite{kavousi2023blindperm}) for data privacy into DAG-based DLTs could prove advantageous. Techniques such as homomorphic encryption~\cite{vaikuntanathan2011computing}, commitment schemes~\cite{pedersen1991non}, multiparty computation~\cite{zyskind2015decentralizing}, and differential privacy~\cite{hassan2020differential} could further bolster privacy in these systems.

Most DAG-based DLTs have unencrypted plaintext metadata on their ledgers and support pseudonymous addresses. This makes DAG-based DLT an unattractive solution for privacy-required applications. Due to the plaintext data on the ledger of these DLTs, an adversary can track the associated accounts using data analysis tools, hindering the identity of the users and afterward conducting harmful activities to steal or freeze their funds. In this manner, the privacy of users' identities, transaction data, and linkability between transactions are compromised~\cite{wang2020preserving}. In the blockchain domain, these issues have been addressed by employing privacy-preserving techniques or by creating privacy-centric blockchain systems~\cite{zhang2019security,feng2019survey}. Nevertheless, to the best of our knowledge, there is no privacy-centric DAG-based DLT. Therefore, in this paper, we present a brief study of privacy-related properties, possible methods to implement privacy, and the associated challenges in implementing privacy in DAG-based DLTs.

\subsection{DAG-based DLT}
Organizing a distributed ledger in a DAG topology enables higher performance and lower costs compared to the traditional linearly structured blockchains. In a DAG-based DLT~\cite{raikwar2024sok}, transactions are stored in the vertices of a directed graph, and multiple references per vertex are possible. The directed graph contains no directed cycles. 

\begin{definition}{(DAG-based DLT)} A DAG-based DLT is a system in which multiple nodes jointly maintain a ledger of transactions structured as a DAG. An edge between two vertices represents the causal relationship between the corresponding transactions included in these vertices. The final state of each node corresponds to a set of non-conflicting transactions from vertices in various branches of the DAG.
\end{definition}

In a linear blockchain, the longest chain rule serves a dual purpose: agreeing on the included blocks and establishing their order. Nevertheless, the DAG structure enables massively parallel transaction validation and a causal order relation among the transactions (represented by the edges). However, the complex structure of DAG brings unique challenges compared to the traditional blockchains while implementing privacy.



\subsection{Privacy Notions}
There are three main privacy notions in a DLT. To define these notions (informally), let $A_i$ denote an account address of a user $i$, and $tx$ denote a transaction.

\begin{definition} (Confidentiality)
Confidentiality of a transaction $tx$ holds if the content (payload) of the transaction $tx$ issued by an account $A_i$ is indistinguishable from the content of any other transaction $tx'$ either issued by $A_i$ or by any other account $A_j$, i.e., transaction data is not public to the users.
\end{definition}



\begin{definition} (Anonymity)
   Anonymity of a user $i$ holds if account $A_i$ of the user cannot be linked to transactions $tx_1, \ldots, tx_m$ which are produced by $A_i$.  
\end{definition}

Anonymity is also referred to as Untraceability.

\begin{definition} (Unlinkability)
    Unlinkability holds if two transactions $tx, tx'$ are produced by an account $A_i$, or by different related accounts, it is impossible to sufficiently distinguish whether the transactions $tx, tx'$ are related or not.
\end{definition}

In a recent work~\cite{PDAG}, Wang et al. proposed a privacy-preserving transaction DAG (PDAG) which captures the unlinkability, untraceability, and source concealment of the transactions. However, it does not construct a privacy-preserving DAG-based DLT. Therefore, in the current landscape of DAG-based DLT, the solutions and the associated challenges to achieve privacy notions have not been explored. 

Achieving these privacy notions brings significant advantages for the DAG-based DLTs and their users. Specifically, ordering the transactions in a DAG-based DLT equipped with privacy counters Miner Extractable Value (MEV) attacks~\cite{daian2020flash}. These MEV attacks aim to manipulate the order of client transactions for personal gain. However, by concealing transaction details through privacy measures, adversaries face greater difficulty in executing MEV attacks. This way, not only does privacy provide MEV protection but it also gives fairness in transaction ordering.



\subsection{Contribution}
To the best of our knowledge, no prior work focuses on privacy notions in DAG-based DLTs. In this work, we propose existing cryptographic techniques to achieve privacy on the state-of-the-art DAG-based DLTs. Motivated by the above, contributions of this paper are as follows:

\begin{enumerate}
    \item We present a list of privacy challenges that occur in every DLT (Section~\ref{sec:privacy-challenges}) and in DAG-based DLTs (Section~\ref{sec:privacy-DAG-DLTs}).
    \item We briefly describe the solutions to achieve privacy along with associated challenges and instances. (Section~\ref{sec:privacy-solutions}).
    \item We present a few interesting future research directions (Section~\ref{sec:future-directions}).
\end{enumerate}




\section{Privacy Challenges in DLTs} \label{sec:privacy-challenges}

Implementing privacy in DLTs introduces challenges, some of the generic challenges are described as follows:

\subsection{Pseudonymity Vs Anonymity}
One of the main concerns when it comes to privacy in DLT is the issue of pseudonymity and anonymity. Pseudonymity refers to the use of a pseudonym or alias instead of a real name or identity, however, anonymity refers to the state of hiding the real identity, meaning unidentifiable or untraceable. 

In the context of pseudonymity, the identities are obscured by pseudonyms while the transactions are recorded publicly on the ledger. This provides a baseline level of privacy but does not guarantee full anonymity. From the publicly available transaction analysis in DLTs, the pseudonyms can be traced and real identities can be exploited. In DLTs, achieving complete anonymity is more challenging than pseudonymity. 

Anonymity provides strong protection by making the transactions untraceable. To achieve anonymity, privacy-enhancing cryptographic techniques such as ring signature~\cite{Monero}, and Zero-knowledge proofs~\cite{Zcash} can be employed in DLTs. However, it is essential to recognize the limitations and potential vulnerabilities associated with each privacy technique given the specific use case and threat model.

\subsection{Complexity} 
In DLTs, four types of complexity are paramount: communication, computation, verification, and storage.

Communication complexity quantifies the information exchange among honest nodes crucial for consensus establishment. This involves factors such as the broadcast protocol employed, the number of messages transmitted, and propagation delay. In DAG-based protocols, communication complexity pertains to the total number of bits transmitted by an honest node to validate and order a single transaction. Research endeavors strive to mitigate this complexity by optimizing message propagation and reducing network overhead.

Computation complexity denotes the computation time required for block generation. Many DLTs adopt puzzle-based mechanisms like Proof-of-Work (PoW)~\cite{nakamoto2008bitcoin} and Proof-of-Stake (PoS)~\cite{nguyen2019proof}, demanding substantial computational resources. Implementing privacy features into such block creation or block propagation processes using privacy-preserving mechanisms invariably escalates computational and communication overhead respectively. Furthermore, privacy-enabled transaction or blocks increases the storage complexity.

\subsection{Regulatory Compliance}
Regulatory compliance and privacy are incompatible in most of the cases. In a privacy-enabled DLT, regulatory compliance can be compatible with privacy as long as it is possible for users to prove certain properties regarding the origin of their funds or transactions. In DLTs, cryptographic primitives can help to achieve that. For example, ring signatures~\cite{sun2017ringct} or anonymous communication channels~\cite{kohlweiss2021anonymity} can enable one to prove those properties without revealing the source. 

\subsection{Auditability}
Auditing the transactions while still maintaining the privacy of the transaction amount and transacting parties is a challenging task. Regulatory compliance mandates such as Anti Money Laundering (AML) and Know Your Customer (KYC) necessitate robust auditing mechanisms within DLT systems. In permissioned DLTs implementing privacy, auditing can be implemented in an easier way (e.g., by assigning a fixed auditing committee) than in a permissionless DLT. Auditability of transactions can be achieved by employing cryptographic primitives, e.g., Zero-knowledge proofs~\cite{kang2019fabzk, li2023auditable}, publicly-verifiable oblivious RAM (PVORM) or audit tokens in Solidus~\cite{cecchetti2017solidus} and zkledger~\cite{narula2018zkledger}, commitment and signature schemes~\cite{androulaki2020privacy}.

DLTs may also incorporate auditing committees tasked with verifying transaction data. The committee can employ a secret sharing mechanism~\cite{pedersen1991non} where each member possesses an encrypted secret share of the private transaction data. Once the majority of committee members agree to decrypt the private transaction, they can view the transaction data. 

\subsection{Usability and Adoption}
Implementing privacy features in a way that is user-friendly and does not hinder the user experience is crucial for the adoption of DLTs. Complex privacy settings or interactions can deter users from utilizing the privacy features. There are many state-of-the-art privacy-preserving cryptocurrencies, however, only a few of these, e.g., Zcash, and Monero are popular due to their user-friendly settings and documentation. Furthermore, ensuring that privacy mechanisms are compatible with other systems/standards is vital for the broader adoption of DLTs. 



\section{Privacy in DAG-based DLTs} \label{sec:privacy-DAG-DLTs}
DAG-based DLTs differ with traditional blockchain DLTs in many aspects. The DAG structure, consensus mechanism, scalability requirements, and transaction finality models bring a few unique features and challenges to DAG-based DLTs. Following, we present a few challenges apart from the generic ones described in previous Section~\ref{sec:privacy-challenges}.





\subsection{In Tip Selection} Tip selection mechanism can become expensive in terms of computation and time complexity while implementing privacy. The order of expensiveness depends on the implemented cryptographic method as well as the tip selection mechanism. Nevertheless, irrespective of the privacy mechanism, the tip selection process checks for the validity of the tips. Therefore, once a privacy-enabled cryptographic primitive is employed on the tips, the validation of these tips may incur an extra cost. 

A recent work by Yang et al.~\cite{yang2024tip} presents a privacy deanonymization attack targeting the IOTA tip selection method. The authors exploit the reliability of the tip selection process of light nodes from the full IOTA nodes. The tip selection results in establishing a direct link between a light node and an adversarial node which leads to compromising users' privacy. The authors also presented ways to improve the privacy of IOTA light nodes against deanonymization attacks.

\subsection{In Conflict Resolution}
In DAG-based DLTs, conflicts between transactions/blocks require resolution using conflict resolution mechanisms to establish the ordering and dependencies among conflicting transactions/blocks. Here, the conflict resolution mechanism resolves conflicts based on some parameters in the DLT and ordering refers to the way the blocks/transactions are organized. Ordering in DAG-based DLTs can be categorized into two main types: partial order and total order. Partial order involves arranging DAG blocks through topological sorting, while total order establishes a linear sequence of blocks. 

Privacy implementations offer significant benefits to ordering within DAG-based DLTs, particularly in countering MEV attacks~\cite{daian2020flash}. However, implementing privacy might create a challenge for conflict resolution if a parameter used for resolving the conflict is also anonymized.

\subsection{In Performance}
Privacy mechanisms often introduce additional computational and storage overhead, which can counteract the scalability benefits of DAG-based DLTs. Recent efforts in DAG-based DLTs focus on reducing the communication complexity and achieving optimal complexity. Nevertheless, the introduction of privacy measures will invariably increase communication and computation complexity. This can lead to the degradation of performance in DAG-based DLTs when implementing privacy.

Therefore, an analysis of the performance of different cryptographic primitives to achieve privacy in DAG-based DLTs can help to find a balance between maintaining high performance and ensuring robust privacy in DAG-based DLTs. 

\subsection{In Consensus}
Secure and efficient management of cryptographic keys for privacy-preserving mechanisms adds another layer of complexity to the consensus protocol. Moreover, the privacy-preserving mechanism can increase the time required for nodes to reach consensus, impacting the system's overall throughput and latency.

The confirmation rule and the finality rules used to confirm and finalize the transactions may be impacted within the consensus protocol due to the implemented privacy in the respective DAG-based DLTs.


\subsection{In Anonymization}
In the realm of privacy within DAG-based DLTs employing the UTXO transaction model, several key privacy concerns arise which focus on de-anonymize the users and data, such as: address reuse, taint analysis, transaction analysis, and metadata analysis. These concerns were examined by Tennant~\cite{tennant2017improving}, who conducted theoretical analyses on privacy issues within IOTA.

Address reuse entails employing the same address for multiple transactions, potentially compromising privacy. To mitigate this, it is recommended to utilize fresh addresses for each transaction and ensure addresses remain unlinkable to personally identifiable information, thereby enhancing privacy.

Taint analysis seeks to establish associations between pairs of addresses by analyzing transaction graphs. This process involves quantifying the percentage of tokens at a particular address that may have originated from another address, shedding light on transactional relationships.

Transaction analysis delves into the structure of transactions, particularly focusing on input and output addresses. Notably, funds at input addresses are entirely utilized in transactions, with any remaining balance sent to a new address provided by the sender. The identification of change addresses in transactions can potentially compromise privacy.

Metadata analysis concentrates on scrutinizing transaction metadata, which can be exploited to breach privacy. For instance, certain DAG-based DLTs rely on manual peer discovery, necessitating static IP addresses, thereby complicating anonymity and potentially undermining privacy measures.

\subsection{In Attack Vectors}
Implementing privacy can introduce several other attack vectors than de-anonymization attacks in the DAG-based DLTs. A Sybil attack can be mounted where an adversary can create a large number of pseudonymous identities. With the Sybil attack method defined in~\cite{bissias2014sybil}, the real identities of the users can be exploited in DAG-based DLTs.

An adversary can also mount a DoS attack in privacy-enabled DAG-based DLTs, which will increase the computation and verification complexity for the new blocks created/received by the users. An adversary can also exploit side-channel sensitive information from the implemented privacy-preserving technique in the DAG-based DLTs. Furthermore, an adversary can also launch a replay attack by replaying the private (encrypted) transaction in different contexts.



\section{Privacy Solutions} \label{sec:privacy-solutions}
This section outlines privacy implementation mechanisms in DLTs. While these mechanisms have been extensively utilized in blockchain DLTs, their application in DAG-based DLTs remains underexplored. Consequently, this section provides a comprehensive overview of these methods, delineating the privacy notions they encapsulate. Subsequently, we examine the challenges they pose and explore potential benefits and applications in existing and prospective DAG-based DLT contexts.

\subsection{Transaction Mixing} 
Mixing services facilitate the transfer of funds to a new address in an unlinkable manner, offering enhanced privacy for users. When Alice utilizes a mixing service to transfer funds, the service randomly selects funds from another user and transfers them to Alice's designated fresh address. This process makes it indistinguishable to external observers, such as Bob, whether the transaction is a mixing transaction or a regular transaction on the ledger. Alice, as a participant in the mix, can plausibly deny her involvement in the transaction, thereby bolstering privacy protection.

There are two types of mixing techniques: 1) Centralized and 2) Decentralized. Centralized mixing requires trust assumption on the mixing service. A corrupt centralized mixing service can do malicious activities such as stealing the users' funds during mixing. Even if the centralized mixing service is not corrupt, it could be attacked or forced to reveal the mixing information of users' funds, hence losing the privacy of user transaction data. 
Nevertheless, decentralized mixers solve these problems. However, depending on the DAG-based DLT design and requirements, any of the two techniques can be employed to achieve unlinkability.

\textbf{Privacy Notions} Mixing services are promising solutions for unlinkability property. Nevertheless, mixing services can also be used to preserve anonymity property. Confidentiality is not achieved by general mixing services.

\textbf{Challenges} Mixing services obscure users' transactions but these services are also exploited by money launderers or Sybil attacks. Therefore, a risk assessment is required to carefully perform the identity checks (e.g., KYC) of the transacting users in such a way that the unlinkability property among transactions in the mixers is still preserved.

\textbf{Instances} There have been a few works implementing privacy in DAG-based DLTs by employing the mixing technique. Obyte platform (formally byteball~\cite{Blackbytes}) offers built-in smart contract payment using its privacy currency Blackbytes which can include privacy features such as mixing and blinding. However, in Blackbytes, the payment histories were found to be traceable. Later, Tennant presented centralized mixers as an ideal solution for improving anonymity in IOTA 1.0~\cite{popov2018tangle}. Nevertheless, centralized mixers do not even provide anonymity against weaker passive adversaries. Therefore, Sarfaraz et al.~\cite{sarfraz2019privacyIota}, presented a privacy-aware IOTA ledger that uses decentralized mixers and provides unlinkability property for users' transactions of IOTA 1.0~\cite{popov2018tangle}. The authors used a combination of digital multi-signature scheme~\cite{iota-multisig} and decryption mix-nets~\cite{ruffing2014coinshuffle}.  Werner et al.~\cite{werner2020anonymization} implemented privacy in Nano~\cite{lemahieu2018nano} by using centrally administered coinmixers. A similar technique can be employed in Vite~\cite{liu2018vite} which is an extension of Nano to provide the total order of all the users' transactions while anonymizing the users' transactions.


\subsection{Zero-knowledge Proofs}
Zero-knowledge proofs provide a method where a party proves to another party that a given statement is true, without giving any other information than the statement.

The idea behind Zero-knowledge proofs is to allow one party, the prover, to produce concise proof to convince the verifier that the “prover” is performing only correct computations on its private data. Importantly, this technique reveals nothing about the “prover’s” personal data to the verifier

   A basic Zero-knowledge proof system $\mathsf{ZKP}$ involves a prover $P$ and a verifier $V$. Let $R$ be an efficient computable binary relation that has a pair of variables $(x,w)$ where $x$ is a statement and $w$ is a witness. Let $\mathcal{L}$ be the language that defined by $R$, i.e. $\mathcal{L}=\{x |\; \exists \; w\; s.t.\; R(x,w)=1 \}$. A Zero-knowledge proof for language $\mathcal{L}$ is a pair of prover $P$ and verifier $V$ where the prover wants to convince the verifier that $x\in \mathcal{L}$ without revealing $w$. 
   
   \begin{definition}
      A Zero-knowledge proof system consists of a tuple of algorithms:
    \begin{itemize}
        \item $\mathsf{Setup}(1^{\lambda})$: Given a security parameter $\lambda$, output a common reference string (CRS) $crs$.
        \item $\mathsf{Prove}(x,crs,w)$: Given a statement $x$, $crs$, and witness $w$, output a proof $\pi$ .
        \item $\mathsf{Verify}(x,crs,\pi)$: Given a statement $x$, $crs$, and proof $\pi$, output accept or reject.
    \end{itemize}
       \end{definition} 

Zero-knowledge proofs have been explored for the blockchains~\cite{Zcash, Pribank} with the motivation of providing privacy. However, later these proofs were employed in blockchains for scalability purposes. For the DAG-based DLTs, the interest in Zero-knowledge proof is recently growing. 



\textbf{Privacy Notions} Zero-knowledge proofs can be used to achieve all privacy notions. Zero-knowledge proofs are widely used to provide confidentiality of smart contract data or off-chain transaction data. It has also been used to provide anonymity and unlinkability in blockchains, e.g.,~\cite{Zcash}.

\textbf{Challenges} Zero-knowledge proofs are expensive in terms of complexity. In terms of proof size, zk-SNARK can help but from the perspective of computation, it is still complex. However, recent advancement shows an increasing interest and progress in developing efficient Zero-knowledge proofs. However, in DAG-based DLTs, depending upon the areas where these proofs are employed, Zero-knowledge proofs can help to achieve privacy incurring extra costs.

\textbf{Instances} Aleph zero~\cite{Alephzero} system uses Zero-knowledge proofs to implement (level 1) privacy while executing Aleph-BFT DAG-based consensus protocol. Another work related to Zero-knowledge proofs on DAG-based protocols is Snark on Aleo which uses Zero-knowledge virtual machines for executing the Narwhal-Bullshark~\cite{SnarkOSALeo}. Hashgraph~\cite{baird2020hashgraph} employs Zero-knowledge proofs to create identity tokens that can be used to verify the users without revealing personal information about the users. Guardian~\cite{norkin2022guardian}, a policy engine, creates a marketplace of tokenized data assets on Hashgraph and uses a selective disclosure mechanism by employing Zero-knowledge proofs.

Zero-knowledge proofs have also been experimented~\cite{greitbauer2023exploring} to create a relay contract in IOTA~\cite{popov2020coordicide} which provides validation of milestones created by the coordinator.


\subsection{Digital Signature} 

Digital signatures are inevitable cryptographic primitives in the blockchain. The concept of digital signature came forward from public-key cryptography. Digital signatures are used to verify the authenticity of digital messages and documents. Signatures are widely used in different applications such as contract signings, software distribution, and financial transactions.

Signatures are a fundamental building block in blockchains. Signatures are primarily used to verify the authenticity of blockchain transactions. By providing a signature over a transaction, a user proves that he/she is authorized to spend the funds of the transaction while preventing other users from spending those funds.

\begin{definition} A signature scheme consists of three algorithms:
\begin{itemize}

    \item $\mathsf{KeyGen}(1^{\lambda})$: Given a security parameter $\lambda$, output public-private key-pair $(pk,sk)$.
    \item $\mathsf{Sign}(M,sk)$: Given a message $M$ and a secret key $sk$, output a signature $\sigma$.
    \item $\mathsf{Verify}(M,\sigma,pk)$: Given a message $M$, signature $\sigma$, and a public key $pk$, output accept or reject.
\end{itemize}
\end{definition}

Digital signatures are widely used in DLT space for signing blocks and transactions. However, there are different types of signature schemes that can be used to provide privacy: 1) Ring signature~\cite{rivest2001} where the idea is to create a signature on a message on behalf of a spontaneous group of signers while preserving the identity of the signer. Ring signatures are used to provide anonymity;
2) Blind signature~\cite{Chaum1984} where the idea is to disguise the message before signing. The message is blinded by combining it with some blinding factor.

\textbf{Privacy Notions} Depending upon the type of signature, different privacy notions are achieved. Blind signatures are used to provide confidentiality while ring signatures are used to provide anonymity. Unlinkable ring signatures can be used to provide the notion of unlinkability.

\textbf{Challenges} A general ring signature size is linear to the participants in the ring (anonymity set). Therefore, once there are many participants in the DLT, the signature size could increase, resulting in an increase in storage complexity. A constant-size ring signature can be used to avoid the issue of signature size. While using blind signatures, conflict resolution on the blinded transactions could be challenging. 

\textbf{Instances} There has been no research on the applicability of ring signatures in DAG-based DLT. A ring signature can be relatively easily employed in DAG-based DLTs which operate in a permissioned network than in a permissionless network. Examples of DAG-based DLTs where ring signature can achieve anonymity of users include Hashgraph~\cite{baird2020hashgraph}, Jointgraph~\cite{xiang2021jointgraph}, Caper~\cite{amiri2019caper}. CDAG~\cite{gupta2019cdag} can also use a ring signature for the elected participants in each slot. Furthermore, the blind signature can be applied in most of the DAG-based DLTs, e.g., Fino~\cite{malkhi2022maximal}.

\subsection{Encryption} 
Encryption is a mechanism to provide confidentiality of data. It is a process of converting a representation of information (plaintext) into another representation (ciphertext) so that only authorized parties can access the original information from the ciphertext representation. In an encryption scheme, a sender $S$ encrypts a plaintext using a key and sends the output ciphertext to receiver $R$, then the receiver decrypts the ciphertext using a key.

In the blockchain, encryption plays an important building block to achieve privacy of data. Encryption schemes are used in blockchain to hide transaction data such as transaction value or asset information. Encryption is also used to provide privacy for users' account balances. 

\begin{definition}
An encryption scheme involves three algorithms:
\begin{itemize}
    \item $\mathsf{KeyGen}(\lambda$): Given a security parameter $\lambda$, output an encryption key $ek$ and a decryption $dk$.
    \item $\mathsf{Enc}(M, ek$): Given a plaintext $M$ and an encryption key $ek$, output a ciphertext $C$.
    \item $\mathsf{Dec}(C, dk$): Given a ciphertext $C$ and a decryption key $dk$, output a message $M$.
\end{itemize}
\end{definition}


\textbf{Privacy Notions} Encryption provides confidentiality property for transaction data.

\textbf{Challenges} Encryption of conflicting transactions makes the conflict resolution process hard. Moreover, achieving total order on the encrypted transactions can also be a challenge depending on the metric for ordering in the respective DAG-based DLT. The encryption of transactions allows blind-order fairness for the respective DLT. The other notion of fairness might become hard to achieve depending on the complexity of the encryption method.

\textbf{Instances} In DAG-based DLTs, encryption techniques have been mostly used to provide secure communication between participants, e.g., Hashgraph~\cite{baird2020hashgraph}, Byteball~\cite{churyumov2016byteball}. Fino~\cite{malkhi2022maximal} ledger encrypts the mempool transactions to provide the privacy of transactions with the main goal of MEV resistance. 

Encryption techniques can be applied to provide privacy of transaction data. For example, in DAG-based DLTs where a transaction necessitates the creation of two blocks in the DAG, one for the sender and another for the receiver, encryption can be used by encrypting the transaction using the receiver's public key. This includes DAG-based DLTs, e.g., Nano~\cite{lemahieu2018nano}, Vite~\cite{liu2018vite}, DLattice~\cite{zhou2019dlattice} can provide transaction privacy using encryption. Another example could be Caper~\cite{amiri2019caper}, where cross-application transactions can be encrypted, preserving the privacy of transaction data. 

\subsection{Homomorphic Encryption}
Homomorphic encryption (HE) enables parties to perform simple arithmetic operations, i.e., addition and multiplication, on encrypted data without compromising confidentiality.  
    
    \begin{definition}A homomorphic encryption scheme~\cite{vaikuntanathan2011computing} is a tuple of probabilistic polynomial time algorithms:
    \begin{itemize}
        \item $\mathsf{Setup}(1^{\lambda})$: Given a security parameter $\lambda$, Output global parameters $param$s.
        \item $\mathsf{KeyGen}(params) $: Given global parameters $param$, output a public-private key-pair $(pk,sk)$. 
        \item $\mathsf{Enc}(params, pk,\mu)$: Given $param$, $pk$, and a message $\mu \in R_\mathcal{M}$, output a ciphertext $c$. 
        \item $\mathsf{Dec}(params, sk, c)$: Given $param$, $sk$, and a ciphertext $c$, output a message $\mu^* \in R_\mathcal{M}$.
        \item $\mathsf{Eval}(pk, f , c_1 , . . . , c_l )$: Given the public key $pk$, a function $f : R_\mathcal{M}^l \rightarrow R_\mathcal{M}$ which is an arithmetic circuit over $R_\mathcal{M}$, and a set of $l$ ciphertexts $c_1 , . . . , c_l $, output a ciphertext $c_f$. 
    \end{itemize}
    
    \end{definition}
    
    In the above scheme, the message space $\mathcal{M}$ of the encryption schemes is a ring $R_\mathcal{M}$, and the functions to be evaluated are represented as arithmetic circuits over this ring, composed of addition and multiplication gates.

    Homomorphic encryption can be of different types depending upon the types of arithmetic operations allowed and a limit on the number of operations.

    Once deemed limited, the field of HE is rising and becoming mainstream similar to ZKP. Not only academia but also industries are exploring the advancement of HE and its application. Microsoft, Intel, and DARPA have launched a program to accelerate the development of Fully Homomorphic Encryption (FHE). Moreover, the Zama industry has enabled private smart contracts using FHE. Zama claims that ``{FHE smart contracts are doable today, with a throughput of ~5 tps. FHE ASICs will enable 1,000+ tps at a fraction of the cost~\cite{Zama}.}'' In their private smart contract implementation, users of the blockchain encrypt data using FHE, and the arithmetic operations are performed following the FHE scheme without the need for decryption. This also prevents the possibility of MEV attacks. In addition, the suggested interplay by Zama, between ZKP for scalability and FHE for privacy in DLTs is an interesting research area for further exploration.

    There have been a few works in DLTs using blockDAG structure that employ homomorphic encryption to provide privacy on transaction data~\cite{Dero, Xelis}. A similar technique can be applied in other DAG-based DLTs to incorporate transaction (data) privacy in their systems. Xelis~\cite{Xelis} uses Partially Homomorphic Encryption (PHE) using Elgamal cryptosystem~\cite{elgamal1985public}.

\textbf{Privacy Notions} Homomorphic encryption preserves the confidentiality property. 
    
\textbf{Challenges} The current DLTs with blockDAG structure use the PHE scheme. Employing a full FHE is still considered slow, and therefore, not used in DAG-based DLTs as it might disrupt the performance (scalability) of DAG-based DLTs. The majority of FHE schemes have a growing size of ciphertext due to additional noise vectors, creating a challenge to be employed in DLTs. Nevertheless, finding a suitable HE scheme for a DAG-based DLT application is a challenging task and needs further investigation.
 
\textbf{Instances} Homomorphic encryption can be applied in most of the DAG-based DLTs, considering the DLT doesn't sacrifice scalability over the computational complexity of HE. DAG-based DLTs supporting smart contracts can readily utilize HE to perform computation over encrypted inputs.

\subsection{Other Methods}

\subsubsection{Anonymous Broadcast} In a DLT, the identities of the users/miners are revealed during the broadcast even when the transaction data has been anonymized using a privacy-preserving mechanism, e.g., ZKP. The identities can be anonymized by making the broadcast anonymous. This can be achieved using anonymous broadcast channels~\cite{kotzanikolaou2017broadcast} or by the use of Tor~\cite{ishai2006cryptography}. 

\textbf{Privacy Notions} Anonymous broadcast captures anonymity and unlinkability notions of privacy. 

\textbf{Challenges} Implementing anonymous broadcast channels requires careful consideration of leakage of any metadata. Even though these have been suggested in PoS blockchains to preserve the identity of the stakeholder, Kohlweiss et al.~\cite{kohlweiss2021anonymity} proved that even an ideal anonymous broadcast channel can not preserve the identity of stakeholders. Moreover, a DoS attack might disconnect a Tor node from the DLT network~\cite{bitcoin-tor}. Therefore, a careful investigation is needed in order to use anonymous broadcasts in DAG-based DLTs.

\textbf{Instances} MACT, a multi-channel anonymous consensus mechanism~\cite{li2023mact} uses anonymous broadcast based on Tor to protect node privacy. The ledger of MACT is structured as a DAG. The anonymous broadcast method could be employed to provide confidentiality in cross-chain transactions of DAG-based DLTs, e.g., Nano~\cite{lemahieu2018nano}, Caper~\cite{amiri2019caper}, CDAG~\cite{gupta2019cdag}. 

\subsubsection{Trusted Execution Environment (TEE)} It provides a hardware-protected secure and isolated execution environment for code and data within the computing node. Hence, it provides confidentiality and integrity of code and data. The popular implementations of TEE are Intel SGX~\cite{costan2016intel} and ARM Trustzone~\cite{pinto2019demystifying}. TEE can be used by the participating nodes of DAG-based DLT for encrypting the transaction data, secure signing and verification of blocks, or ordering the blocks by the timestamps stored within the TEE. With the use of TEE, MEV attacks within DAG-based DLT can be prevented and fairness in the ordering can be enforced. 

\textbf{Privacy Notions} TEE preserves the confidentiality of data.

\textbf{Challenges} Use of TEE within DAG-based DLT brings some challenges. First, all the participating nodes must have TEE in them. Second, TEE might affect the performance. Therefore, research is needed on how to incorporate TEE in DAG-based DLTs while not sacrificing performance.

\textbf{Instances} TEE has been employed in DAG-based systems during the block construction and ordering protocols, e.g., Teegraph~\cite{fu2022teegraph} and TEEDAG~\cite{lu2023teedag}. TEE can be used in permissioned DAG-based DLTs, e.g., Tusk~\cite{danezis2022narwhal}, Bullshark~\cite{spiegelman2022bullshark}, Fino~\cite{malkhi2022maximal} for transaction privacy (resulting in order-fairness).

    
    
    
    




\begin{table*}[!htb]
    
    \caption{Summary of Privacy Solutions}
    \centering
    
    \begin{tabular}{|l|l|l|c|c|}
      \hline
      & & & & \\
      \textbf{Privacy Scheme} & \textbf{\makecell{Privacy Notions\\ Captured}} & \textbf{\makecell{Challenges}} & \textbf{Existing Instances} & \textbf{Possible Instances} \\[0.2cm] \hline 
      \textbf{Transaction Mixing} & \makecell[l]{Unlinkability \\ Anonymity} & \makecell[l]{In attack vectors \\ In anonymization} & \cite{sarfraz2019privacyIota, werner2020anonymization} & Vite~\cite{liu2018vite} \\[0.2cm] \hline

      \textbf{Zero-knowledge Proofs} & \makecell[l]{All notions} & \makecell[l]{In complexity \\ In tip selection} & \cite{Alephzero, SnarkOSALeo, norkin2022guardian} & \makecell{Jointgraph~\cite{xiang2021jointgraph}, Tusk~\cite{danezis2022narwhal}, \\Bullshark~\cite{spiegelman2022bullshark}}\\[0.2cm]   \hline

      \textbf{Digital Signature} & \makecell[l]{Depends on the \\ signature scheme} & \makecell[l]{In complexity \\ In conflict resolution} & \cite{malkhi2022maximal} & \makecell{Hashgraph~\cite{baird2020hashgraph}, Caper~\cite{amiri2019caper},\\ Jointgraph~\cite{xiang2021jointgraph}, CDAG~\cite{gupta2019cdag}}\\[0.2cm] \hline

      \textbf{Encryption} & Confidentiality & In conflict resolution & \cite{baird2020hashgraph, churyumov2016byteball, malkhi2022maximal} & \makecell{Nano~\cite{lemahieu2018nano}, Vite~\cite{liu2018vite}, \\ DLattice~\cite{zhou2019dlattice}, Caper~\cite{amiri2019caper}} \\[0.2cm] \hline

      \textbf{Homomorphic Encryption} & Confidentiality & In performance & -- & -- \\[0.2cm] \hline

       \textbf{Anonymous Broadcast} &  \makecell[l]{Unlinkability \\ Anonymity} & \makecell[l]{In attack vectors \\ In anonymization} & \cite{li2023mact} & \makecell{Nano~\cite{lemahieu2018nano}, Vite~\cite{liu2018vite}, \\ CDAG~\cite{gupta2019cdag}, Caper~\cite{amiri2019caper}} \\[0.2cm] \hline

       \textbf{Trusted Execution Environment} & Confidentiality & In performance & \cite{lu2023teedag, fu2022teegraph} & \makecell{Tusk~\cite{danezis2022narwhal}, Bullshark~\cite{spiegelman2022bullshark}, \\ Fino~\cite{malkhi2022maximal}}\\[0.2cm] 
        
      \hline
    \end{tabular}
    \label{tab:comparison}
    \end{table*}

\section{Current landscape and Future Research} \label{sec:future-directions}
The current landscape of DAG-based DLT does not prioritize privacy. Nevertheless, for the broader adoption of DAG-based DLTs, it is vital to provide some notions of privacy that users of these DLTs can choose. Though all of the described cryptographic techniques to achieve privacy in DAG-based DLTs as shown in Table~\ref{tab:comparison} are used in blockchain, only a few of these techniques are actually being tested for DAG-based DLTs to achieve privacy. Therefore, there is a dire need for research on the applicability of these techniques in DAG-based DLT space to find robust solutions to achieve privacy without sacrificing the scalability goal of these DLTs.

Future research should focus on the integration of privacy techniques and the trade-off between privacy and performance. For example, the current implementations of the technique as ZKP require significant computational capacity to construct zero-knowledge proofs. With the increase in the complexity of the problem statement, the proof generation may involve executing arithmetic circuits in the range of ${10}^6$ to ${10}^{12}$, which is computationally intensive. Furthermore, ZKP requires better hardware as in Graphic Processing Units (GPUs) that have parallel processing capability to reduce the resources and time needed to prove a statement. Therefore, research is needed that can combine ZKP with other privacy-preserving techniques, e.g., signatures or encryption which can reduce the complexity and further won't degrade the performance. 


\section{Conclusion} \label{sec:conclusion}
In this paper, we conducted a comprehensive examination of privacy considerations within the framework of DAG-based DLTs. Our investigation began with an exploration of the overarching privacy challenges inherent in DLTs and subsequently into DAG-based DLTs. Moreover, we briefly explained the existing mechanisms available for incorporating privacy within DAG-based DLTs. Within the scope of each mechanism, we scrutinized the applicability and potential utilization of privacy measures in various DAG-based DLT contexts. Further exploration of the outlined privacy mechanisms could involve their integration into established DAG-based DLT systems. Additionally, researchers may select and implement these mechanisms to architect a privacy-focused DAG-based DLT tailored to specific requirements.

\section{Acknowledgement}
The author has been supported by IOTA Ecosystem Development grant.


\bibliographystyle{IEEEtran}
\bibliography{report}

\end{document}